\documentclass[letter]{aa}  
\usepackage{graphicx}
\usepackage{natbib}
\bibpunct{(}{)}{;}{a}{}{,}
\usepackage[Symbol]{upgreek}
\usepackage{hyperref}
\usepackage{url}
\usepackage{txfonts}
\usepackage{stfloats}
\usepackage{amstext}
\begin{document} 

   \title{The Close AGN Reference Survey (CARS):}
   \subtitle{Discovery of a global [\ion{C}{ii}]~158$\upmu$m line excess in AGN HE~1353$-$1917}
   \author{I. Smirnova-Pinchukova\inst{\ref{mpia}}
          \and B. Husemann\inst{\ref{mpia}}
          \and G. Busch\inst{\ref{uni_koeln}}
          \and P. Appleton\inst{\ref{caltech}}
          \and M. Bethermin\inst{\ref{amu}}
          \and F.~Combes\inst{\ref{lerma}}
          \and S.~Croom\inst{\ref{uni_sydney}}
          \and T.~A.~Davis\inst{\ref{cardiff}}
          \and C. Fischer\inst{\ref{dsi}}
          \and M. Gaspari\inst{\ref{uni_princeton}}\fnmsep\thanks{\textit{Lyman Spitzer Jr.} Fellow}
          \and B. Groves\inst{\ref{uni_australia}}
          \and R. Klein\inst{\ref{sofia-usra}}
          \and C.~P.~O'Dea\inst{\ref{uni_manitoba},\ref{rochester}}
          \and M.~Pérez-Torres \inst{\ref{granada},\ref{zaragoza}}
          \and J.~Scharw\"achter\inst{\ref{gemini}}
          \and M.~Singha\inst{\ref{uni_manitoba}}
          \and G.~R.~Tremblay\inst{\ref{CfA}} 
          \and T.~Urrutia \inst{\ref{aip}}
          }

   \institute{Max-Planck-Institut f\"ur Astronomie, K\"onigstuhl 17, 69117 Heidelberg,                                          Germany \email{smirnova@mpia.de} \label{mpia} 
              \and I. Physikalisches Institut der Universit\"at zu K\"oln, Z\"uplicher Str. 77, 50937 K\"oln, Germany \label{uni_koeln}
              \and Caltech/IPAC, 1200 E. California Blvd., Pasadena, CA 91125, USA \label{caltech}
              \and Aix Marseille Univ, CNRS, CNES, LAM, Marseille, France \label{amu}
              \and LERMA, Observatoire de Paris, PSL Research Univ., Coll\`ege de France, CNRS, Sorbonne Univ., UPMC, Paris, France \label{lerma}
              \and Sydney Institute for Astronomy, School of Physics, A28, The University of Sydney, NSW, 2006, Australia \label{uni_sydney}
              \and School of Physics \& Astronomy, Cardiff University, Queens Buildings, The Parade, Cardiff, CF24 3AA, UK \label{cardiff}
              \and Deutsches SOFIA Institut, Pfaffenwaldring 29, 70569 Stuttgart, Germany\label{dsi}
              \and Department of Astrophysical Sciences, Princeton University, 4 Ivy Lane, Princeton, NJ 08544-1001, USA \label{uni_princeton}
              \and Research School of Astronomy and Astrophysics, The Australian National University, Cotter Road, Weston, ACT 2611, Australia\label{uni_australia}
              \and SOFIA-USRA, NASA Ames Research Center, MS 232-12, Moffett Field, CA 94035, USA\label{sofia-usra}
              \and Department of Physics \& Astronomy, University of Manitoba, Winnipeg, MB R3T 2N2, Canada \label{uni_manitoba}
              \and School of Physics \& Astronomy, Rochester Institute of Technology, 84 Lomb Memorial Drive, Rochester, NY 14623, USA \label{rochester}
              \and Instituto de Astrofísica de Andaluc\'{i}a, Glorieta de las Astronom\'{i}a s/n, 18008 Granada, Spain \label{granada}
              \and Departamento de F\'{\i}sica Te\'orica, Facultad de Ciencias, Universidad de Zaragoza, E-50009 Zaragoza, Spain \label{zaragoza}
              \and Gemini Observatory, Northern Operations Center, 670 N. A’ohoku Pl., Hilo, Hawaii, 96720, USA \label{gemini}
              \and Center for Astrophysics $|$ Harvard \& Smithsonian, 60 Garden St., Cambridge, MA 02138, USA \label{CfA}
              \and Leibniz-Institut f\"ur Astrophysik Potsdam, An der Sternwarte 16, 14482 Potsdam, Germany \label{aip}
             }

  \date{Submitted March 29, 2019; Accepted May 2, 2019}
  \abstract
  {
  The [\ion{C}{ii}]$\lambda$158$\upmu$m line is one of the strongest far-infrared (FIR) lines and an important coolant in the interstellar medium of galaxies that is accessible out to high redshifts. The excitation of [\ion{C}{ii}] is complex and can best be studied in detail at low redshifts. Here we report the discovery of the highest global [\ion{C}{ii}] excess with respect to the FIR luminosity in the nearby AGN host galaxy HE~1353$-$1917. This galaxy is exceptional among a sample of five targets because the AGN ionization cone and radio jet directly intercept the cold galactic disk. As a consequence, a massive multiphase gas outflow on kiloparsec scales is embedded in an extended narrow-line region. Because HE~1353$-$1917 is distinguished by these special properties from our four bright AGN, we propose that a global [\ion{C}{ii}] excess in AGN host galaxies could be a direct signature of a multiphase AGN-driven outflow with a high mass-loading factor.
  }


   \maketitle

\section{Introduction}
\label{sec:intro}

\begin{figure*}[t]
    \centering
        \includegraphics[width=0.9\linewidth]{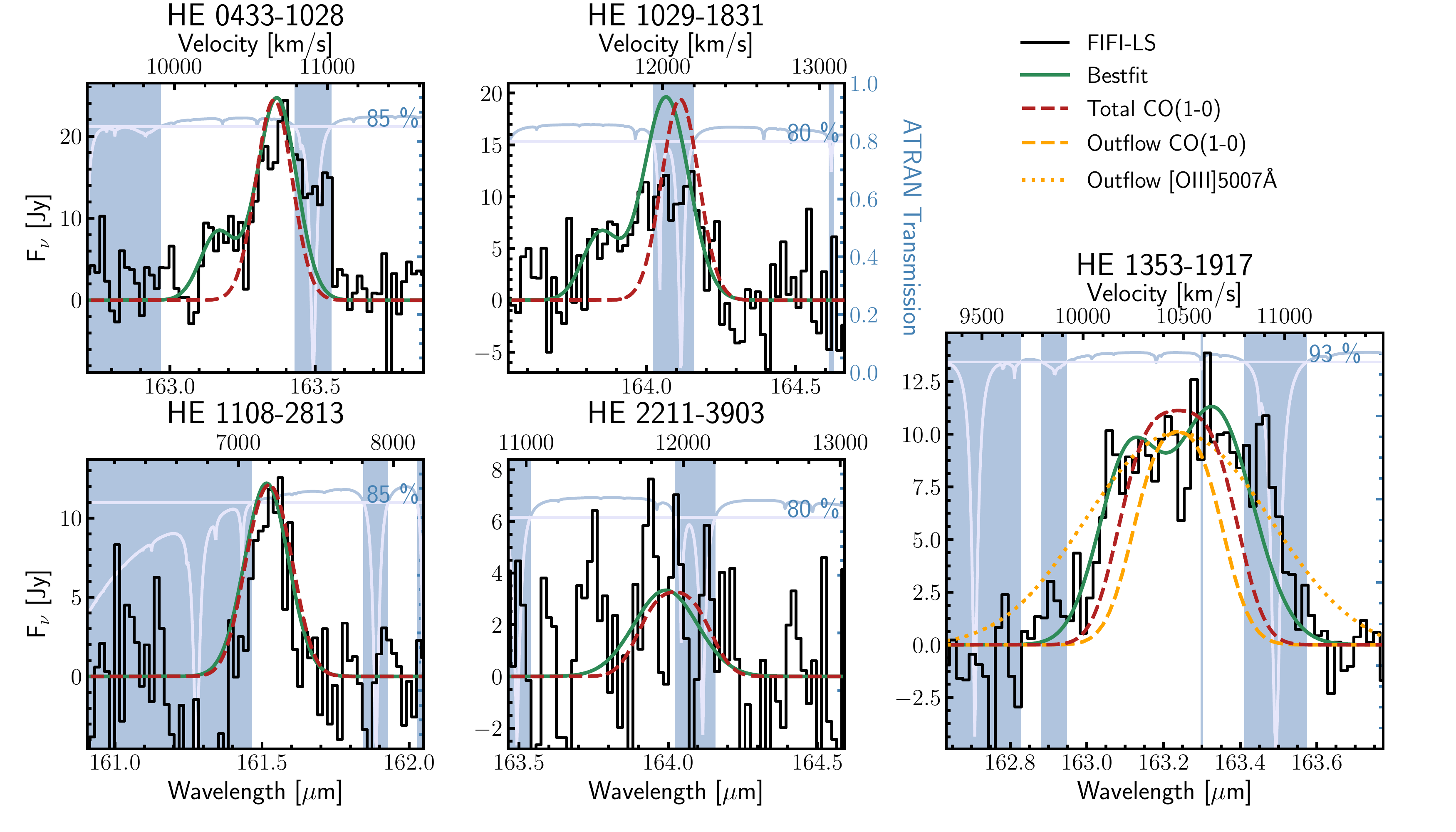}
    \caption{SOFIA/FIFI-LS [\ion{C}{ii}] spectra (black lines) for each object integrated within $36''$ diameter apertures. The CO(1--0) and best-fit Gaussian or double-Gaussian profiles are shown as red dashed and solid green lines, respectively. The CO(1--0) and [\ion{O}{iii}]5007\AA~line profiles of the HE~1353$-$1917 outflow are shown as orange dashed and dotted lines. The comparison line profiles are degraded to the spectral resolution of SOFIA. The atmospheric transmission curves are shown in blue, and the shaded regions of low transmission are excluded from the analysis.}
    \label{fig:spectra}
\end{figure*}

The [\ion{C}{ii}]~157.74\,$\upmu$m emission line arises from the fine-structure transition $^2$P$_{3/2}\rightarrow\,^2$P$_{1/2}$ of the ground state of singly ionized carbon C$^+$ (ionization potential of 11.2\,eV). Working as a coolant in multiple phases of the interstellar medium (ISM), the [\ion{C}{ii}]~line is one of the brightest emission lines in the far-infrared (FIR); it contributes  0.1--0.3\,\% to the FIR luminosity.

The [\ion{C}{ii}]~line has been calibrated as a probe for the cold gas content and associated star formation rates (SFR) in galaxies \citep{stacey1991, boselli2002, herrera-camus2015_sfrcii}. However, using the [\ion{C}{ii}] line as an SFR tracer is complex because of the multiple mixed excitation mechanisms of the line. In local star-forming galaxies, 66--82\,\% of [\ion{C}{ii}] arises from the neutral gas of photodissociation regions (PDRs), and the rest comes from the ionized phase \citep{croxall2017}. The mechanism of dust infrared emission, on the other hand, is rather simple: dust preferentially absorbs UV radiation from the stellar population and therefore is sensitive to the bright young stars. The infrared luminosity has been well calibrated as an SFR tracer at SFR\,$>$1\,$M_\odot\,\mathrm{yr}^{-1}$ \citep{hirashita2003, murphy2011} so that the [\ion{C}{ii}] and FIR luminosity are expected to be correlated.

At the highest SFRs, luminous and ultra-luminous infrared galaxies (U/LIRGs) exhibit the so-called [\ion{C}{ii}] line deficit \citep{helou2001, malhotra2001, luhman2003}, where [\ion{C}{ii}] becomes unreliable as an SFR indicator \citep{diaz-santos2013}. The origin of the line deficit is still debated and is directly connected to physical processes that are crucial for understanding the [\ion{C}{ii}] excitation mechanisms.

Active galactic nuclei (AGN) are able to affect the [\ion{C}{ii}]/FIR ratio in several ways: they can increase the infrared luminosity through dust heating \citep{herrera-camus2018_shining1}; act as an additional source of the [\ion{C}{ii}] excitation; or suppresses the [\ion{C}{ii}] line through the overionization of C$^+$ to C$^{\mathrm{2+,\text{}\,3+,\, etc.}}$ with their hard radiation field \citep{langer_pineda2015}.  \textit{Herschel} surveys of nearby galaxies such as KINGFISH \citep{smith2017} and SHINING \citep{herrera-camus2018_shining2} found no link between the [\ion{C}{ii}] line emission and AGN luminosity, but these AGN may be not luminous enough to outshine the star formation.

Given its brightness, the [\ion{C}{ii}] line is the most important ISM diagnostic at high redshifts that can be observed with unprecedented spatial resolution and depth on submillimeter interferometers. The sample of the observed high-redshift objects includes starburst and AGN-dominated systems, with the [\ion{C}{ii}]/FIR ratios spanning a wide range from 0.02\,\% to 5\,\% \citep[e.g.,][]{gullberg2015, brisbin2015, decarli2018}. In order to provide an interpretation, we need to investigate the [\ion{C}{ii}] emission in local galaxies and determine the effect of luminous AGN.

In this letter, we present [\ion{C}{ii}]~line observations with the Stratospheric Observatory For Infrared Astronomy \citep[SOFIA,][]{temi2014_sofia} for five nearby (0.024\,$<$\,z\,$<$\,0.040) luminous Seyfert\,1 AGN host galaxies from the Close AGN Reference Survey (CARS; \citet{husemann2017_cars}; www.cars-survey.org) to investigate the impact of AGN on the global [\ion{C}{ii}] luminosity.

Throughout the paper, we assume a flat cosmological model with $H_0 = 70$ km s$^{-1}$ Mpc$^{-1}$, $\Omega_M = 0.3$, and $\Omega_{\Lambda} = 0.7$.

\section{Observations and analysis}
\label{sec:obs}

\begin{figure*}[t]
    \centering
        \includegraphics[width=0.95\linewidth]{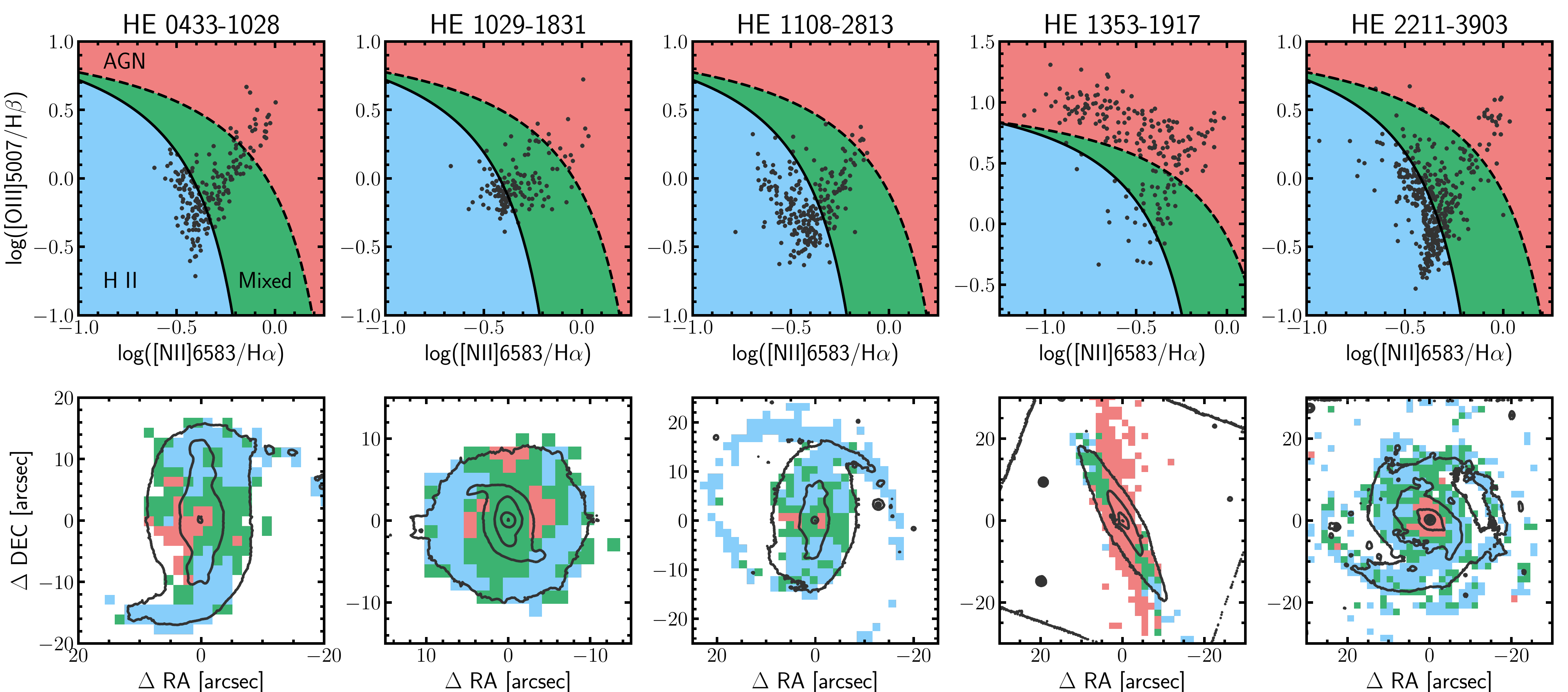}
    \caption{\textit{Upper panel:} MUSE BPT diagnostic diagrams with empirical dividing lines. The solid line shows data from \citet{kauffmann2003_bpt} and the dashed line from \citet{kewley2001_bpt}. The colors represent AGN ionization (red), ionization by \ion{H}{ii} regions (blue), and mixed regime (green). \textit{Lower panel:} MUSE emission line $1.6" \times 1.6"$ binned maps color-coded according to the BPT diagram regions; white-light contours are overplotted in black.}
    \label{fig:bpt}
\end{figure*} 

\subsection{SOFIA/FIFI-LS observations}
\label{sec:fifi}

We observed five CARS objects with the Far Infrared Field-Imaging Line Spectrometer (FIFI-LS; \citet{klein2014_fifi}) on board SOFIA. The objects were picked to cover a broad range of SFRs (1--11\,$M_\odot\,\mathrm{yr}^{-1}$) and avoid the strong atmospheric absorption in the redshifted [\ion{C}{ii}] line wavelength region.  The SFR estimates were initially based on the predictions calculated from the AGN-subtracted extinction-corrected H$\upalpha$ line from observations obtained with the Mulit-Unit Spectroscopic Explorer \citep[MUSE,][section~\ref{sec:muse}]{bacon2010_muse}. The properties of the observed galaxies are listed in Table~\ref{table:obs}.

The observations were performed during SOFIA Cycle~4 (plan ID $04\_0056$, PI:~Husemann) and Cycle~5 (plan ID $05\_0077$, PI:~Husemann). FIFI-LS is a double-beam spectrometer that covers $1'\times1'$ in the red channel (105--200\,$\upmu$m) and $0.5'\times0.5'$ in the blue channel (50--125\,$\upmu$m), split up into $5\times5$ spatial pixels. We tuned the setups to cover the [\ion{C}{ii}] line in the red channel with spectral resolution of R$\sim1200$ (250\,km\,s$^{-1}$) and either [\ion{O}{iii}]\,88\,$\upmu$m or [\ion{O}{i}]\,63\,$\upmu$m in the blue channel, depending on atmospheric transmission.

The pipeline-processed data are provided by the FIFI-LS team. We used LEVEL\_3 science-ready data, which consist of a number of $\sim$30\,second exposures, to apply an additional selection (see Appendix~\ref{sec:time_windows} for details) and background subtraction, and constructed data cubes with $6''$ sampling using the \textsc{Drizzle} algorithm \citep{fruchter1997_drizzle}. To derive total [\ion{C}{ii}] line fluxes from the FIFI-LS cubes, we summed the spectra within an aperture with $36''$ diameter and fit the line shape with Gaussian profiles. The HE~1108$-$2813 and HE~2211$-$3903 spectra are well modeled by a single-Gaussian component, while the spectra of HE~0433$-$1028, HE~1029$-$1831 and HE~1353$-$1917 require two Gaussian components. The spectra and the fitting results together with the excluded wavelength ranges due to the strong atmospheric absorption regions\footnote{Based on the \textsc{ATRAN} tool by Steve Lord, \url{https://atran.arc.nasa.gov/cgi-bin/atran/atran.cgi}} are shown in Fig.~\ref{fig:spectra}. In addition, we have analyzed the blue channel FIFI-LS data for HE~1353$-$1917 to estimate the upper limit of the [\ion{O}{iii}]\,88\,$\upmu$m emission line flux $< 4.8\times10^{-13}$. The [\ion{C}{ii}] best-fit shape was used in the [\ion{O}{iii}] upper limit estimation, taking into account the spectral resolution of R$\sim670$ (450\,km\,s$^{-1}$) for the observed wavelengths.

\subsection{VLT/MUSE observations}
\label{sec:muse}

All five CARS targets were observed with MUSE at the Very Large Telescope (VLT) under ESO programs 094.B-0345(A) and 095.B-0015(A). The MUSE data cover $1\arcmin\times1\arcmin$ FoV at a $0\farcs2$ spatial sampling and a wavelength coverage of  4650--9300\,\AA\ with $R\sim2500$. Integration times range from 400\,s--900\,s split up into two or three exposures, which are rotated by $90^\circ$ against each other for cosmic-ray rejection and better image cosmetics. The data were reduced with the standard MUSE pipeline \citep[version 1.6.0,][]{Weilbacher:2012, Weilbacher:2014}.

As a first step in the post-processing, we subtracted the bright point-like AGN emission from the reconstructed datacube using \textsc{QDeblend$^{\mathrm{3D}}$}, as described in \citet{husemann2013_qdeblend3d, husemann2014_qdeblend3d, husemann2019_he1353}. Afterward we modeled the stellar continuum and ISM emission lines in the AGN-subtracted datacube with \textsc{PyParadise} \citep[see][for details]{husemann2016_pyparadise,weaver2018_pyparadise}.

In Fig.~\ref{fig:bpt} we show the classical Baldwin-Philips-Terlevich \citep[BPT,][]{baldwin1981_bpt} diagrams and corresponding spatial maps for the targeted galaxies after binning 8$\times$8 pixels. Essentially, the BPT diagnostic diagrams highlight different dominating ionization mechanisms. The face-on galaxies display line ratios that are mainly consistent with star formation or intermediate line ratios between star formation and AGN excitation. The edge-on galaxy HE~1353$-$1917 is instead dominated by AGN photoionization. The fraction of AGN excitation to the extinction-corrected H$\upalpha$ for HE~1353$-$1917 is $>$80\%, whereas a fraction of $<$40\% is found for all face-on galaxies.

\begin{table}
\caption{Measured galaxy properties}             
\label{table:obs}      
\centering          
\begin{tabular}{c c c c}
        \hline
    \hline
    Object      &       
    D$_L$       &       
    F$_{[\ion{C}{ii}]}$ &
    F$_{\mathrm{FIR\tablefootmark{*}}}$ \\

        &
    [Mpc]       &       
    $10^{-13}$ [erg/s/cm$^2$] & 
    $10^{-11}$ [erg/s/cm$^2$]  \\
    
    \hline
    HE~0433$-$1028      &       
    156.4       &       
    $6.13 \pm 0.45$    & 
    $9.41 \pm 0.05$ \\
    
    HE~1029$-$1831      &       
    177.7       &       
    $5.32 \pm 0.83$    & 
    $9.82 \pm 0.33$ \\

    HE~1108$-$2813      &       
    104.7       &       
    $2.65 \pm 0.24$    &
    $11.86 \pm 0.32$ \\
    
    HE~1353$-$1917      &       
    154.0       &       
    $5.07 \pm 0.32$    &
    $1.11 \pm 0.04$ \\
    
    HE~2211$-$3903      &       
    175.2       &       
    $1.17 \pm 0.41$    &
    $1.94 \pm 0.25$ \\
    \hline 
\end{tabular}
\tablefoottext{*}{Integrated over 42.5--122.5 $\upmu$m}
\end{table}

\begin{figure*}[t]
    \centering
    \includegraphics[width=0.9\textwidth]{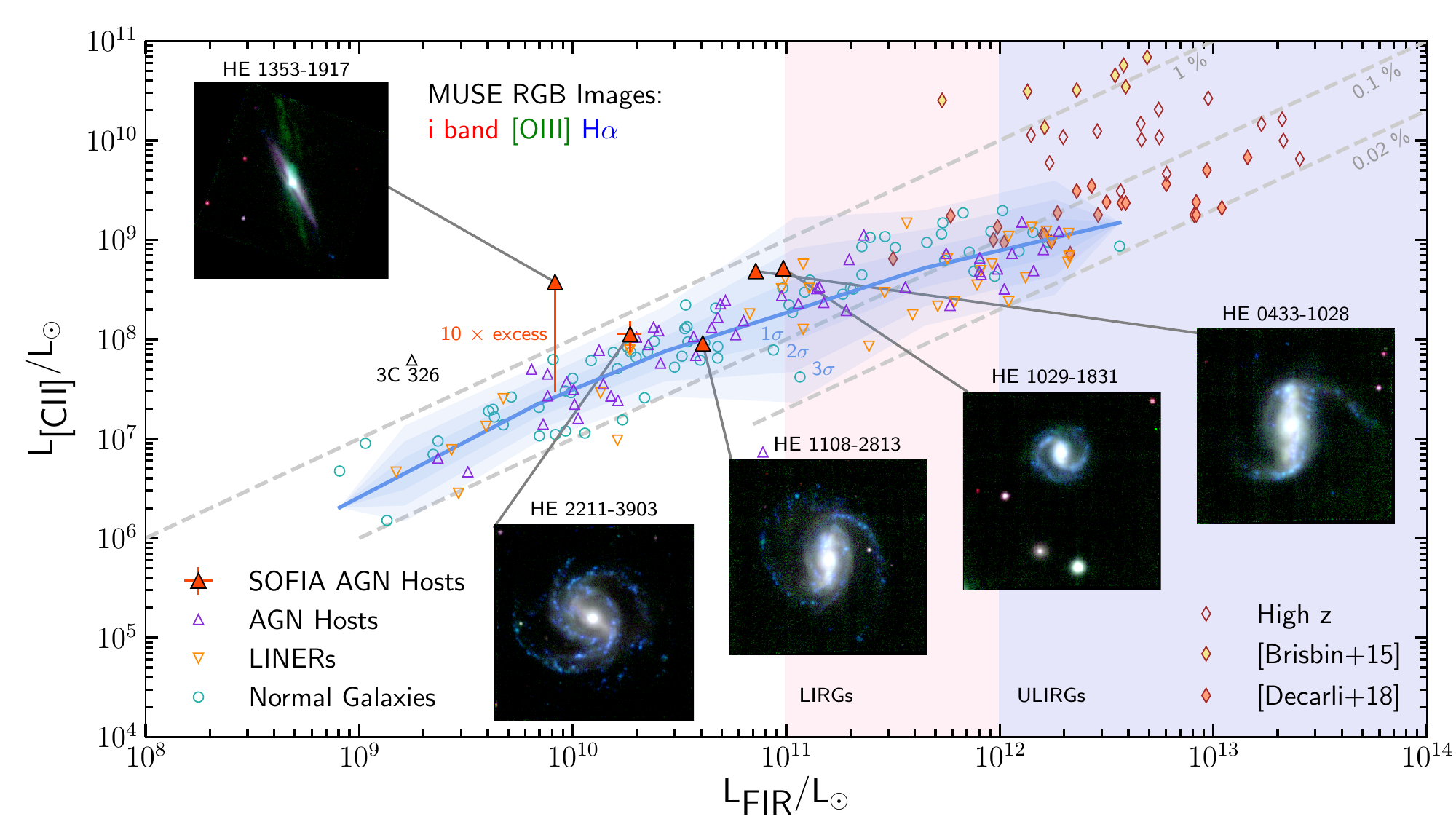}
    \caption{[\ion{C}{ii}] line luminosity as a function of FIR luminosity. Our five CARS targets are shown as red triangles compared to the literature compilation of \citet{herrera-camus2018_shining1}: normal star-forming and star-burst galaxies (green circles), AGN host galaxies (purple triangles), LINER galaxies (orange upside-down triangles), and high-redshift galaxies including the samples from \citet{brisbin2015} and \citet{decarli2018} (brown diamonds). The blue line is the mean of the low-redshift galaxies distribution, and the blue-shaded areas correspond to 1, 2, and 3~$\upsigma$. Pink and purple shades represent LIRG L$_{\mathrm{FIR}} > 10^{11}$ and ULIRG L$_{\mathrm{FIR}} > 10^{12}$ regimes. The luminosities for 3C~326 are taken from \citet{guillard2015}.
    }
    \label{fig:ciivsfir}
\end{figure*}

\subsection{SED fitting and FIR luminosities}
\label{sec:archival_data}
We constructed spectral energy distributions (SEDs) for all targets using publicly available broad-band photometry from the following catalogs: the \textit{Herschel}/SPIRE Point Source Catalogue \citep{schulz2017_spire}, the AKARI/FIS Bright Source Catalogue \citep{yamamura2009_akari_fis}, the 2MASS Extended Source Catalog \citep{jarrett2000_2mass_xcs}, and the GALEX Source Catalog \citep{bianchi2017_galex}. For the following datasets we applied aperture measurements on the survey images: \textit{Herschel}/PACS images from the Herschel Science Archive\footnote{HSA, \url{http://archives.esac.esa.int/hsa/whsa/}}, WISE Image Atlas \citep{cutri2011_wise}, optical \textit{grizy} photometric images\footnote{available at  \url{http://ps1images.stsci.edu}} from the Pan-STARRS DR1 \citep{chambers2016_pan-starrs}, and Swift/UVOT images from the Barbara~A.~Mikulski Archive for Space Telescopes\footnote{MAST, \url{http://archive.stsci.edu/}}.

We obtained the FIR~(42.5--122.5\,$\upmu$m) luminosities through SED fitting with \textsc{AGNfitter} \citep{agnfitter}. \textsc{AGNfitter} uses an MCMC approach and various template libraries to model the SED as a super-position of emission from an AGN accretion disk, a torus of AGN-heated dust, stellar light from the galaxies, and cold dust in star-forming regions. The \textsc{AGNfitter} output SED models are shown in Appendix~\ref{sec:fir_sed} and the FIR~(42.5--122.5\,$\upmu$m) luminosities are listed in the Table~\ref{table:obs}.

\section{Results and discussion}
\label{sec:results}

The [\ion{C}{ii}] and FIR luminosities of our five targets are shown in Fig.~\ref{fig:ciivsfir} in comparison to a large literature compilation of low-redshift galaxies from \citet{herrera-camus2018_shining1}. Four of the CARS targets lie within the 3\,$\upsigma$ region of the mean relation, while HE~1353$-$1917 lies above the relation with more than 7\,$\upsigma$ significance. The outlier has the strongest deviation from the mean trend observed so far at low redshift, with an unprecedented global [\ion{C}{ii}] line excess of an order of magnitude. AGN hosts and LINERs tend to be below the relation in the (U)LIRG regime L$_{\mathrm{FIR}} > 10^{11} \mathrm{L}_{\odot}$ and more likely show [\ion{C}{ii}] line deficits.

Based on the FIR and AGN-subtracted extinction-corrected H$\upalpha$ luminosity, we expected HE~1353$-$1917 to be the faintest target in [\ion{C}{ii}], but it turned out to be the brightest source of our sample. While all galaxies have similar metallicities, stellar masses, and AGN bolometric luminosities, the obvious dissimilarity is the edge-on orientation of HE~1353$-$1917. When we consider the unobscured nature of this AGN, this means that the AGN ionization cone directly pierces the gas-rich disk of the galaxy. This leads to a large biconical extended narrow-line region (ENLR) that is oriented almost along the disk axis on kiloparsec scales, as clearly shown in Fig.~\ref{fig:bpt}. In addition, a massive multiphase outflow on kiloparsec scales with a mass outflow rate of $\dot M_\mathrm{out}\sim$10--100\,$M_\odot\,\mathrm{yr}^{-1}$ is detected in this galaxy, as discussed in detail by Husemann et al. (2019), while less prominent outflows are detected in the other four galaxies (Singha et al. in prep.). The global SFR of $\sim$2$M_\odot\,\mathrm{yr}^{-1}$ implies an integrated mass-loading factor of $\sim$10 or more that has a similar scale as the observed [\ion{C}{ii}] excess.

Several powering sources may contribute to the [\ion{C}{ii}] line excess in HE~1353$-$1917, but the challenge is to distinguish the dominant [\ion{C}{ii}] line excitation mechanism.
As highlighted in Fig.~\ref{fig:ciivsfir}, the [\ion{C}{ii}] line luminosity of HE~1353$-$1917 cannot be powered by the star formation alone. The SFRs calculated from the extinction-corrected H$\upalpha$ ($\mathrm{SFR}_{\mathrm{H}\upalpha}=1.23 \pm 0.03$ $M_\odot\mathrm{yr}^{-1}$) and 42.5--122.5\,$\upmu$m luminosity ($\mathrm{SFR}_{\mathrm{FIR}}=2.3\pm0.1$ $M_\odot\mathrm{yr}^{-1}$) can account for only about 25\% of the observed [\ion{C}{ii}] line luminosity \citep[SFR$_{[\ion{C}{ii}]}=2.286\cdot10^{-43}\times\mathrm{L_{[\ion{C}{ii}]}^{1.034}}$,][]{herrera-camus2015_sfrcii}. X-ray dominated regions (XDRs) produced by the hard X-ray photons from an AGN may contribute and even dominate the PDR [\ion{C}{ii}] emission. From the scaling relation L$_{\mathrm{[\ion{C}{ii}],\,XDR}}=2\times10^{-3}$\,L$_{2-10\mathrm{\,keV}}$ by \citet{stacey2010}, we estimate an XDR contribution of only $10 \%$ given an X-ray luminosity of L$_{2-10\mathrm{\,keV}}=1.69\times10^{43}\mathrm{erg\,s^{-1}}$ (Husemann et al. 2019).

[\ion{C}{ii}] emission can originate in any gas phase that is illuminated by UV photons. Using other line and continuum diagnostics, we can obtain some idea of which phases contribute to the [\ion{C}{ii}] emission, for example, 66--82\,\% for the neutral phase and the rest for the ionized phase in local star-forming galaxies \citep{croxall2017}. However,  these fractions do not necessarily remain the same in the ENLR of an AGN. How much of the [\ion{C}{ii}] emission originates from an ENLR has not been systematically explored so far. If the ionized gas phase produces more than 20--40\,\% of [\ion{C}{ii}] as in PDR paradigm, then it can explain the observed [\ion{C}{ii}] excess. In HE~1353$-$1917 only $\sim$20\,\% of H$\upalpha$ originates from star-forming regions and $\sim$80\,\% comes from the AGN-ionized regions. If we naively assume that 20\,\% of the [\ion{C}{ii}] flux originates from star formation, the [\ion{C}{ii}]$_{\mathrm{SF}}$ of HE~1353$-$1917 falls within 3$\upsigma$ of the [\ion{C}{ii}]--FIR relation.

\begin{figure}[t]
    \includegraphics[width=1.\linewidth]{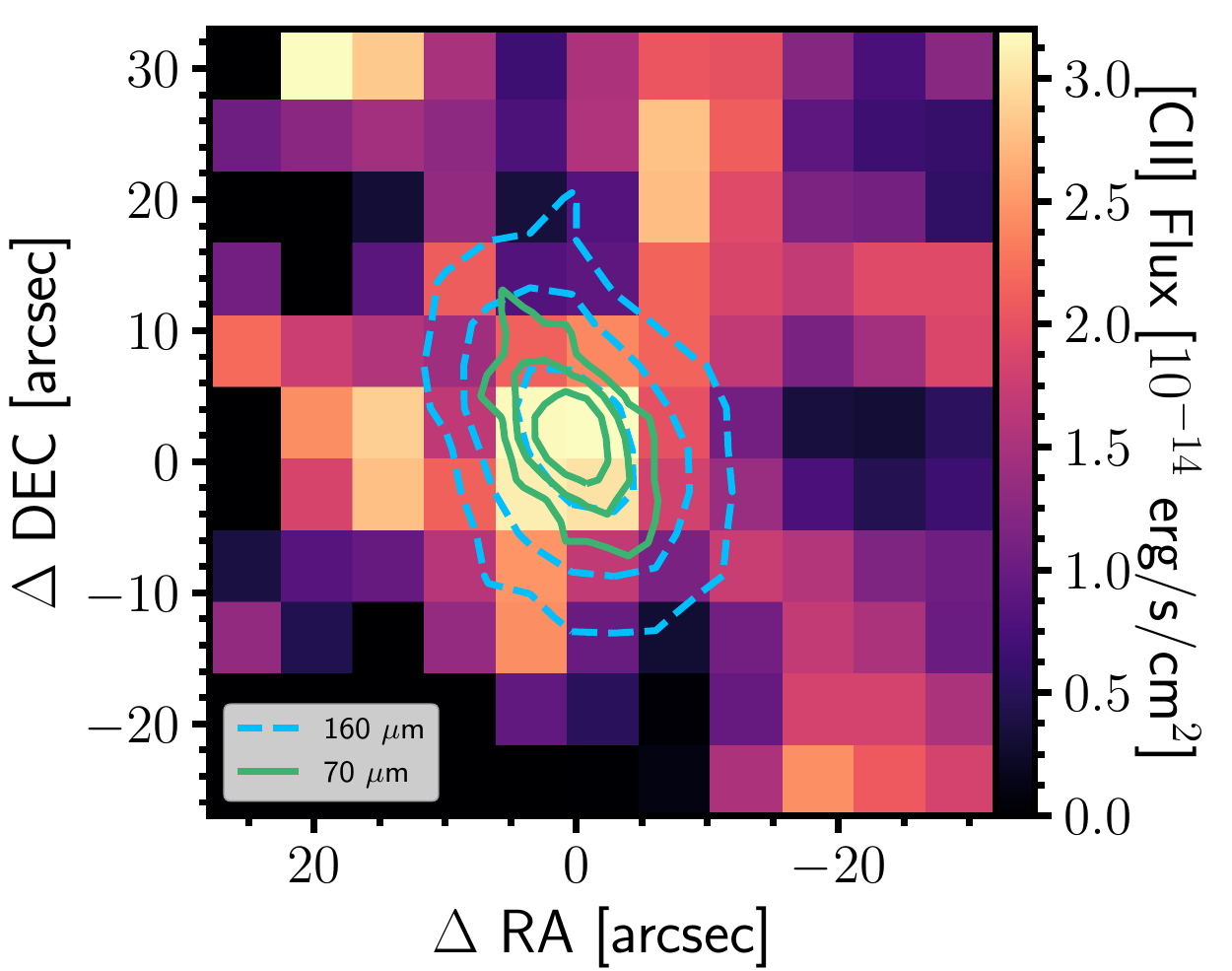}
    \caption{SOFIA/FIFI-LS [\ion{C}{ii}]~line flux map of HE~$1353-1917$.
    The contours represent Herschel PACS 70 (solid green lines) and 160~$\upmu$m (blue dashed line) photometric observations.}
    \label{fig:ciiherschelmap}
\end{figure}

Dissipation of the kinetic energy of shocks and outflows can also have important consequences on the [\ion{C}{ii}] emission, as theoretically explored by \citet{lesaffre2013}. [\ion{C}{ii}]/FIR ratios of 3--7\% are detected in between merging galaxies \citep{appleton2013, peterson2018_taffy} and locally within galaxies \citep{appleton2018}. On the global scale, the radio galaxy 3C~326 emits around $3 \%$ of FIR in [\ion{C}{ii}] line for which jet-driven turbulence is likely responsible \citep{guillard2015}. \citet{nesvadba2010_3C326} estimated an outflow rate in 3C~326 of $\dot M_\mathrm{out}$\,$\sim$\,35\,M$_\odot\,\mathrm{yr}^{-1}$, compared to the SFR$\sim$0.1\,$M_\odot\,\mathrm{yr}^{-1}$ \citep{ogle2007_3C326}, brings a mass-loading factor of $>$10 similar to that found in HE~1353$-$1917.  HE~1353$-$1917 has a global [\ion{C}{ii}]/FIR ratio of $4.3 \pm 0.4 \%$ and a radio-jet powering the multiphase gas outflow (Husemann et al. 2019). The  [\ion{C}{ii}] line width of $760\pm60$\,km\,s$^{-1}$ is broader than the CO(1-0) line for cold molecular gas outflow $\sim$210\,km\,s$^{-1}$, but narrower than the ionized gas outflow $\sim$1020\,km\,s$^{-1}$ (see Fig.~\ref{fig:spectra}). [\ion{C}{ii}] therefore likely traces the interface between the warm and cold gas phase within the outflow. We can relate the region where the [\ion{C}{ii}] excess occurs to the center of the galaxy (Fig.~\ref{fig:ciiherschelmap}) where the brightest region of the ENLR and the kpc-scale multi-phase outflow are indeed located. 
The [\ion{O}{iii}] upper limit implies a line ratio limit of [\ion{O}{iii}]/[\ion{C}{ii}] $< 0.96,$ which suggests that the pure AGN ionization in the ENLR is not the primary cause for the [\ion{C}{ii}] excess.
Hence, the observations are firm evidence that the [\ion{C}{ii}] line excess in HE~1353$-$1917 is related to the multiphase outflow initiated by the jet.

High-redshift galaxies span an entire range between [\ion{C}{ii}] deficient and normal regimes with [\ion{C}{ii}]/FIR ratios from 0.02\,\% to 5\,\%. The large scatter in the ratios may be caused by the order-of-magnitude difference in spatial resolutions, for instance, 18$''$ for CSO, \citet{brisbin2015} with average ratio of 1.8\,\% and 1$''$ for ALMA, \citet{decarli2018} with average ratio of 0.08\,\%, as low-resolution observations can include a larger fraction of intergalactic [\ion{C}{ii}] emission. Several high-redshift QSO have shown a [\ion{C}{ii}] enhancement \citep{maiolino2009, wagg2010}, which was interpreted as due to a low metallicity given the high [\ion{C}{ii}]/FIR and [\ion{C}{ii}]/CO($1-0$) ratios in local dwarf galaxies. We measured [\ion{C}{ii}]/FIR $\sim 4\%$ and [\ion{C}{ii}]/CO($1-0$) $\sim 10^4$ for HE~1353$-$1917, which can neither be explained by intergalactic [\ion{C}{ii}] emission nor by a low metallicity.

\section{Conclusion}
\label{sec:conclusion}
We presented the discovery of a global [\ion{C}{ii}] line excess in one out of five AGN host galaxies. Based on ancillary information from an extensive multiwavelength analysis of this galaxy as part of the CARS survey, we can directly connect the [\ion{C}{ii}] line excess in HE~1353$-$1917 to the impact of the AGN that drives a massive multiphase outflow on kiloparsec scales embedded in an ENLR (Husemann et al. 2019). The detection of such a global [\ion{C}{ii}] excess in AGN host galaxies is of crucial importance for the interpretation of [\ion{C}{ii}] line observations and the detection of massive gas outflows in luminous high-redshift AGN host galaxies. Given the evidence of HE~1353$-$1917 and the similar outflow seen in 3C~326, we propose that a significant [\ion{C}{ii}] line excess in luminous AGN, if detected, can be used as an inference for a multiphase AGN outflow with a high mass-loading factor even at high redshifts. 

\begin{acknowledgements}
      We thank the anonymous referee for helpful comments that improved the quality of the manuscript.
      
      Based on observations made with the NASA/DLR Stratospheric Observatory for Infrared Astronomy (SOFIA). SOFIA is jointly operated by the Universities Space Research Association, Inc. (USRA), under NASA contract NNA17BF53C, and the Deutsches SOFIA Institut (DSI) under DLR contract 50 OK 0901 to the University of Stuttgart.
      
      Based on observations collected at the European Organization for Astronomical Research in the Southern Hemisphere under ESO programme 094.B-0345(A) and 095.B-0015(A).
      
      M.G. is supported by the Lyman Spitzer Jr. Fellowship (Princeton University) and by NASA Chandra GO7-18121X and GO8-19104X.
      
      MPT acknowledges support from the Spanish MINECO through grant AYA2015-63939-C2-1-P.
\end{acknowledgements}

\bibliographystyle{aa}
\bibliography{references.bib}

\begin{appendix}
\section{FIFI-LS time window selection}
\label{sec:time_windows}
\begin{figure}
    HE~0433$-$1028 \\
    \includegraphics[width=1.\linewidth]{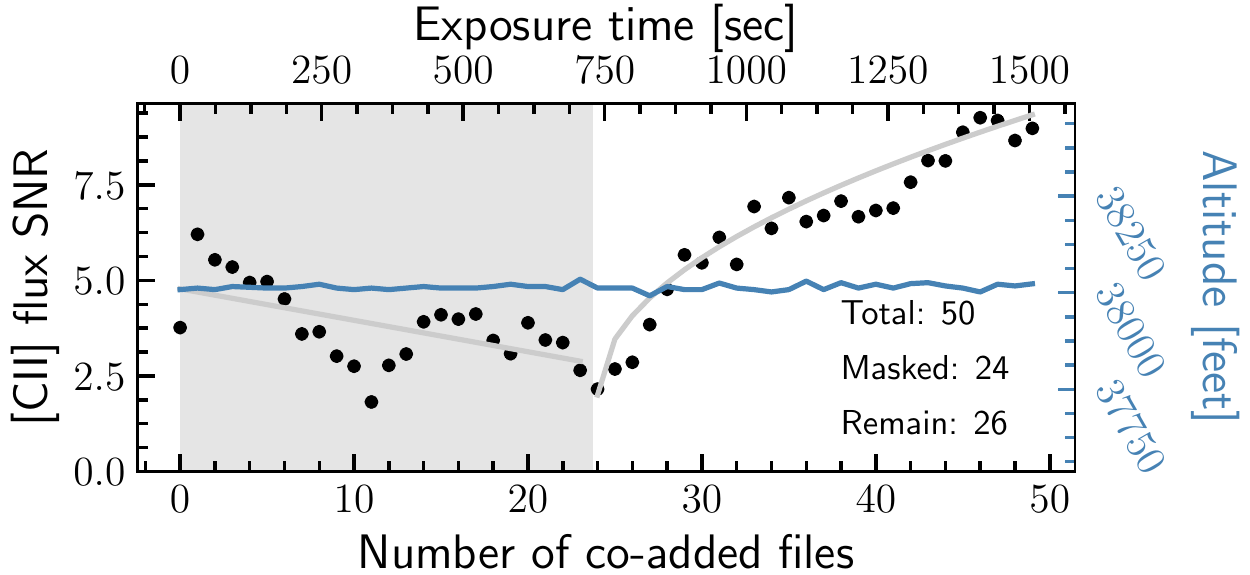}
    HE~1029$-$1831 \\
    \includegraphics[width=1.\linewidth]{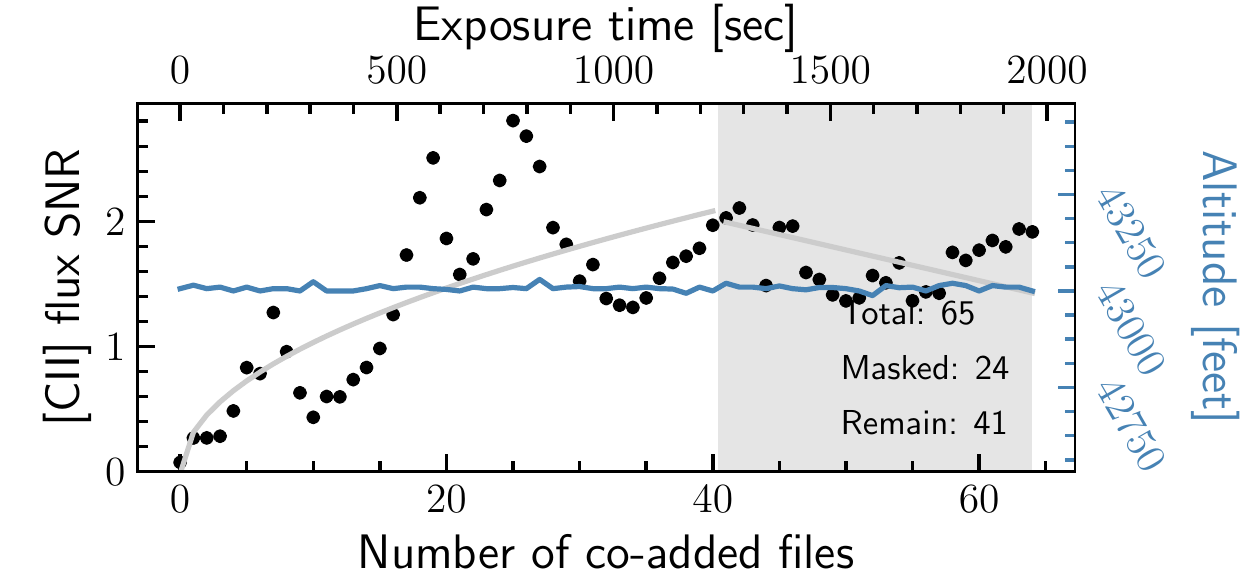}
    HE~1108$-$2813 \\
    \includegraphics[width=1.\linewidth]{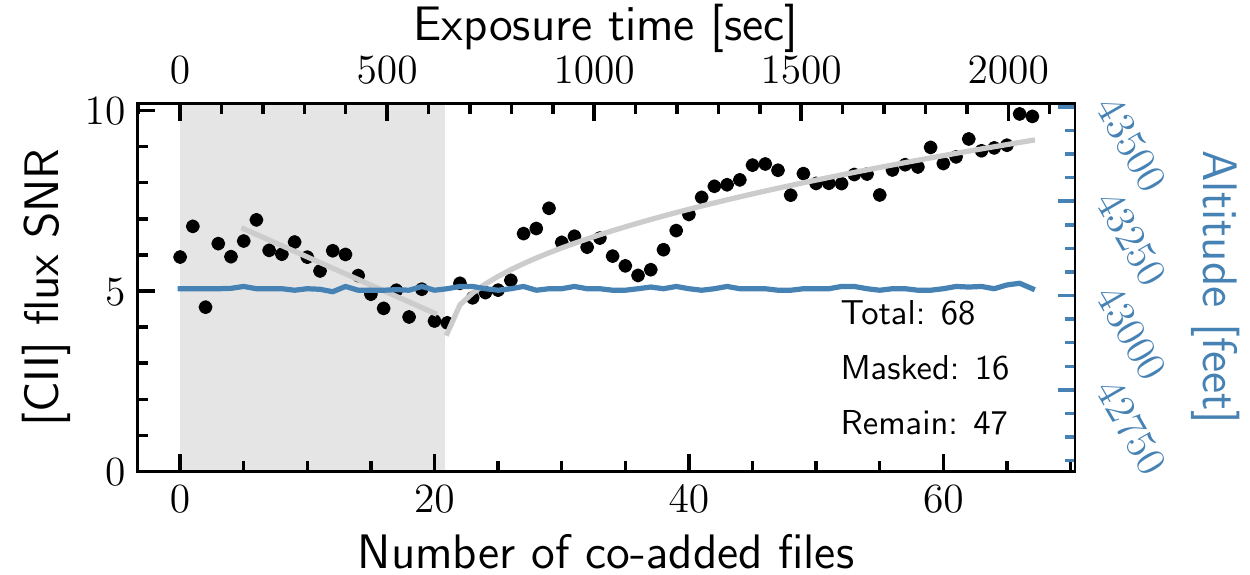}
    HE~1353$-$1917 \\
    \includegraphics[width=1.\linewidth]{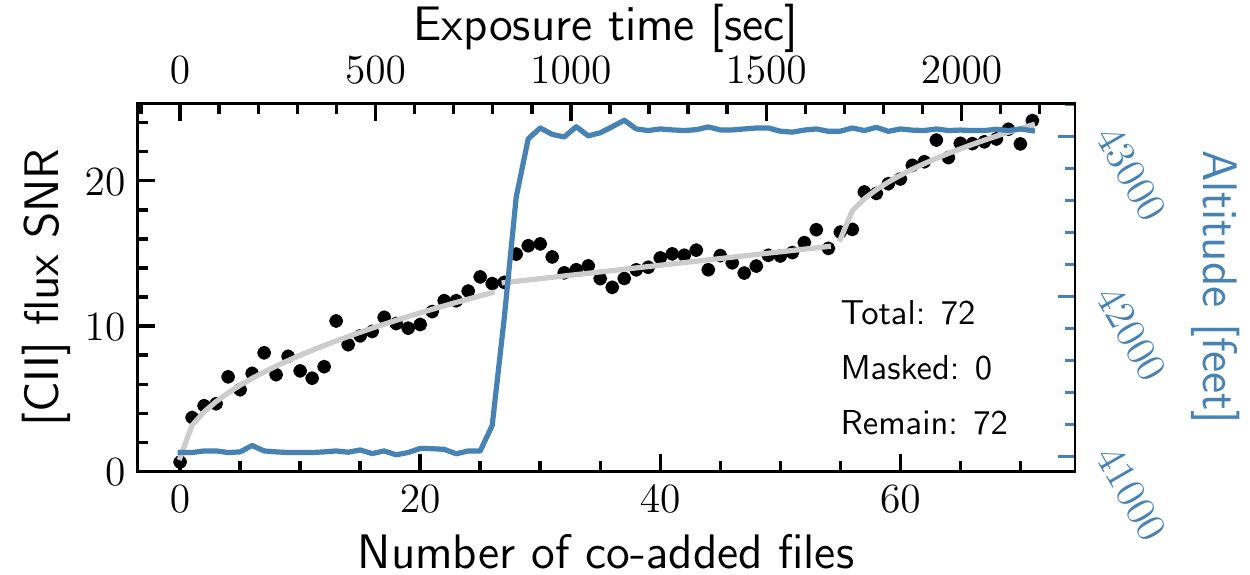}
    HE~2211$-$3903 \\
    \includegraphics[width=1.\linewidth]{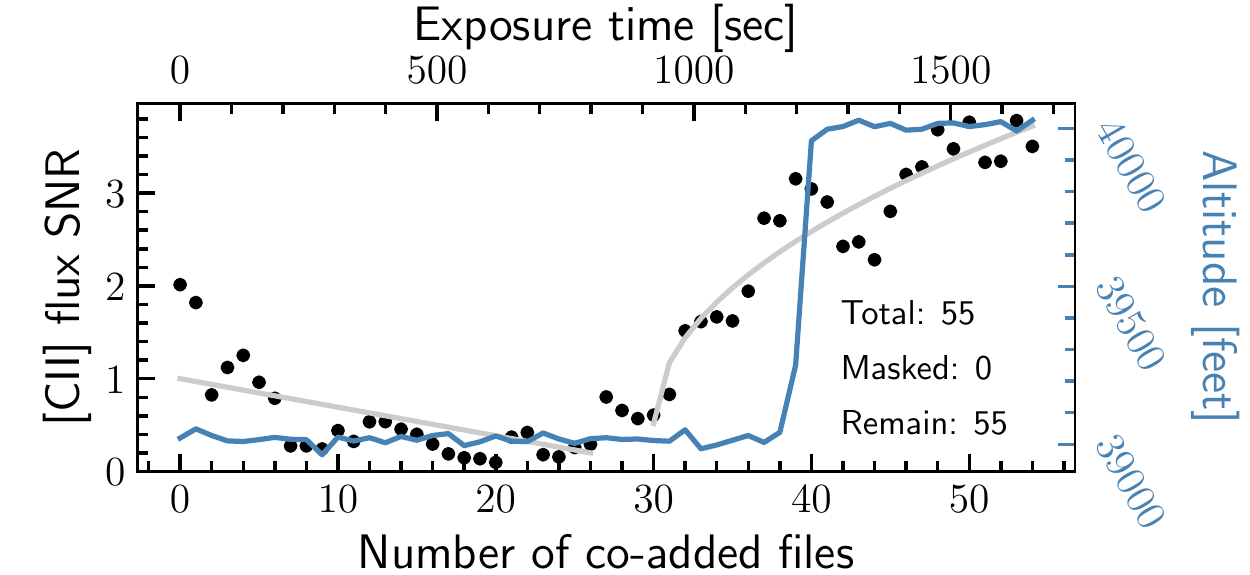}
    \caption{ [\ion{C}{ii}] line S/N as a function of the cumulatively coadded exposures (black circles). The trends (gray lines) are purely subjective and are shown to guide the eye. The shaded areas cover the time windows and associated files that were excluded from the further analysis. The corresponding altitudes of SOFIA are shown as blue lines.}
    \label{fig:exptest}
\end{figure}

Although SOFIA reduces the water vapor absorption to 99\% of the ground level, the atmospheric variations still play a perceptible role and  affect the quality of the FIFI-LS data. To probe this effect, we used the science-ready LEVEL\_3 pipeline output files. Each file corresponds to only $\sim 30$~seconds of exposure time and cannot be examined independently due to the high levels of noise. Calculating [\ion{C}{ii}] line flux signal-to-noise ratio (S/N) of cumulatively summed files allows us to trace the trends of the S/N and therefore judge the atmospheric conditions. Ideally, the trend should rise as the square root function. If the S/N remains the same or even decreases when exposure time is added, we assume that an atmospheric variation degrades the quality and exclude those time windows from the analysis, as shown with the shaded areas in Fig.~\ref{fig:exptest}. 

To calculate the S/N as a function of the exposure time, we fit a fixed Gaussian shape taken from the fitting result before the additional selection to the spectra, summed within the 36$''$ diameter aperture. In the case of HE~1353$-$1917, when the line shape is not Gaussian, the procedure still works similarly because only relative trends are important in this analysis. Masking the spectral regions with strong atmospheric absorption lines is, conversely, very important for the flux estimations. The first few values of the S/N as a function of the coadded files are unreliable, but the overall trends remained when we performed the reverse-order coadding test. 

The rapid altitude shifts also affect the S/N trends as seen in the case of HE~1353$-$1917 and HE~2211$-$3903. In the last 15 files for HE~2211$-$3903 the altitude is higher and of much better quality, but we decided not to exclude the majority of the files from the analysis and kept all of them.

Even though the selection due to atmospheric conditions is performed in the reduction pipeline, our additional selection technique helps to increase S/N of the detection by excluding the poor quality artifact-rich files.

\clearpage
\section{SED fitting}
\label{sec:fir_sed}
\begin{figure}
    \includegraphics[width=1.\linewidth]{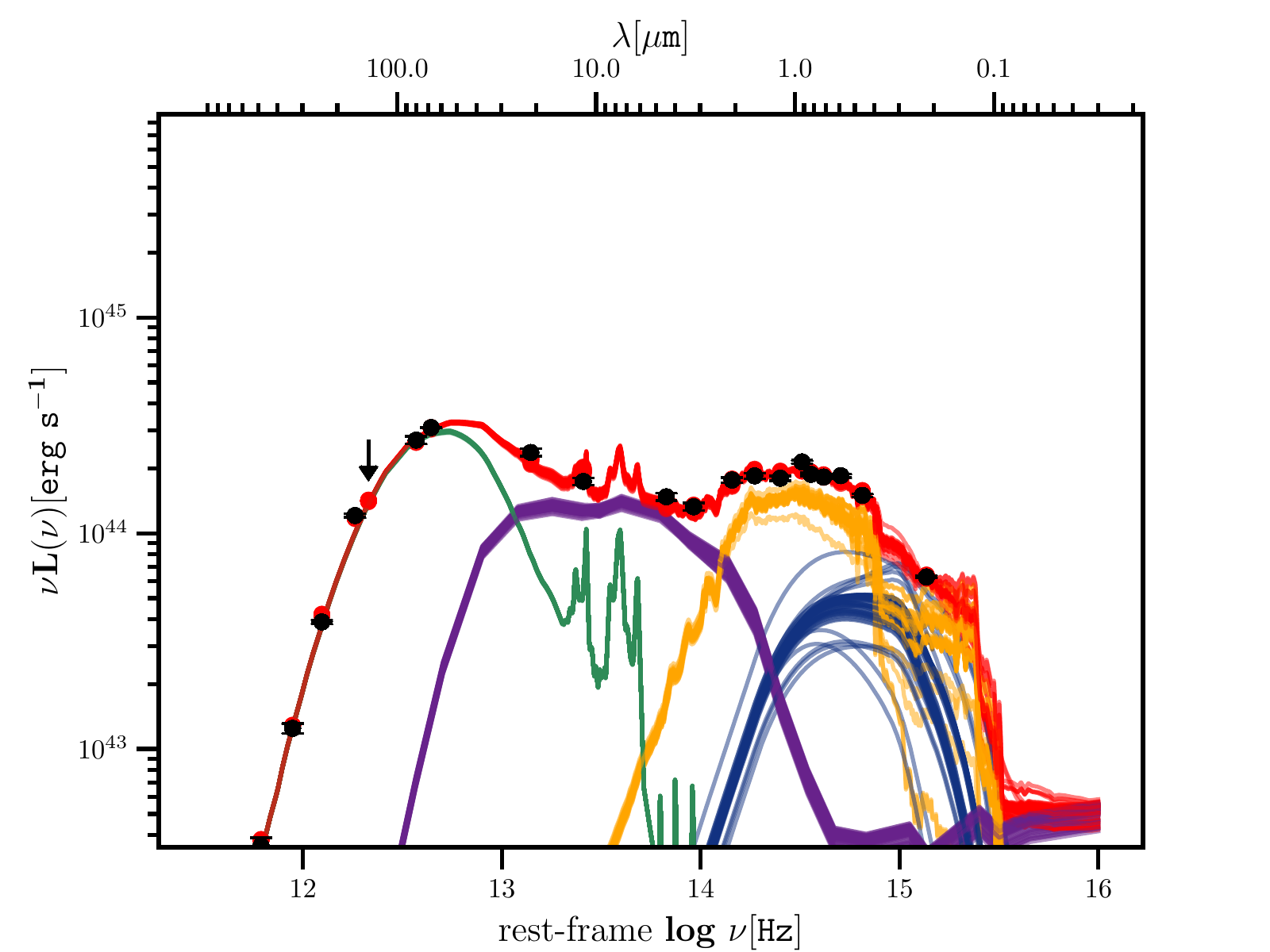}
    \caption{Reconstructed SED for HE~0433$-$1028 and best-fit SED model determined by \textsc{AGNfitter} \citep{agnfitter}. We show 50 MCMC realizations fit to the broadband photometric data (black points with error bars). The red lines represent the total model, which consists of four components: the cold and warm dust in star-forming regions (green lines), the torus of AGN-heated dust (purple lines), the stellar continuum (yellow lines), and the AGN accretion disk (blue lines).}
    \label{fig:SED_HE0433}
\end{figure}

\begin{figure}
    \includegraphics[width=1.\linewidth]{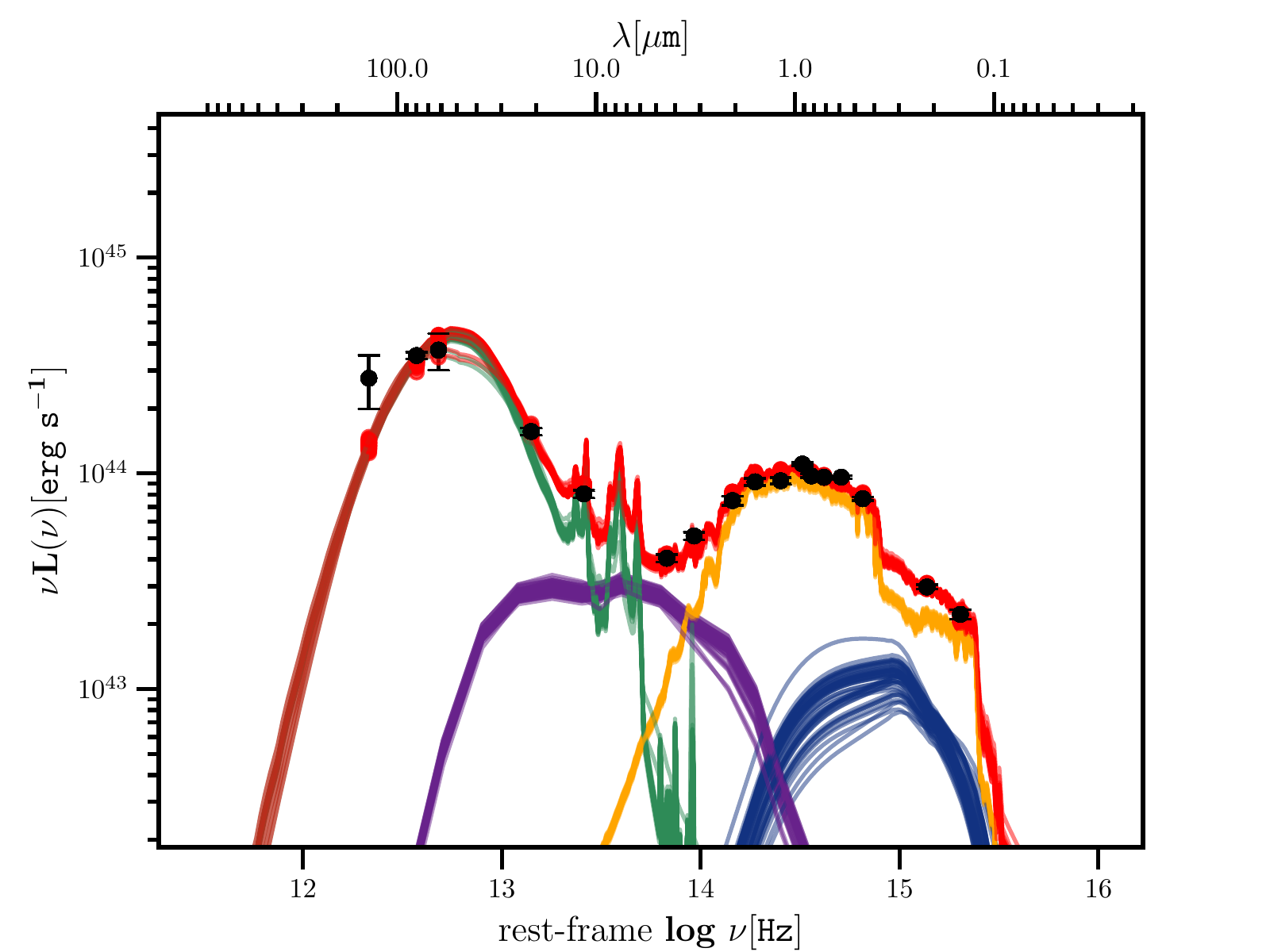}    
    \caption{Same as Fig.~\ref{fig:SED_HE0433} for HE~1029$-$1831.}
    \label{fig:SED_HE1029}
\end{figure}

\begin{figure}
    \includegraphics[width=1.\linewidth]{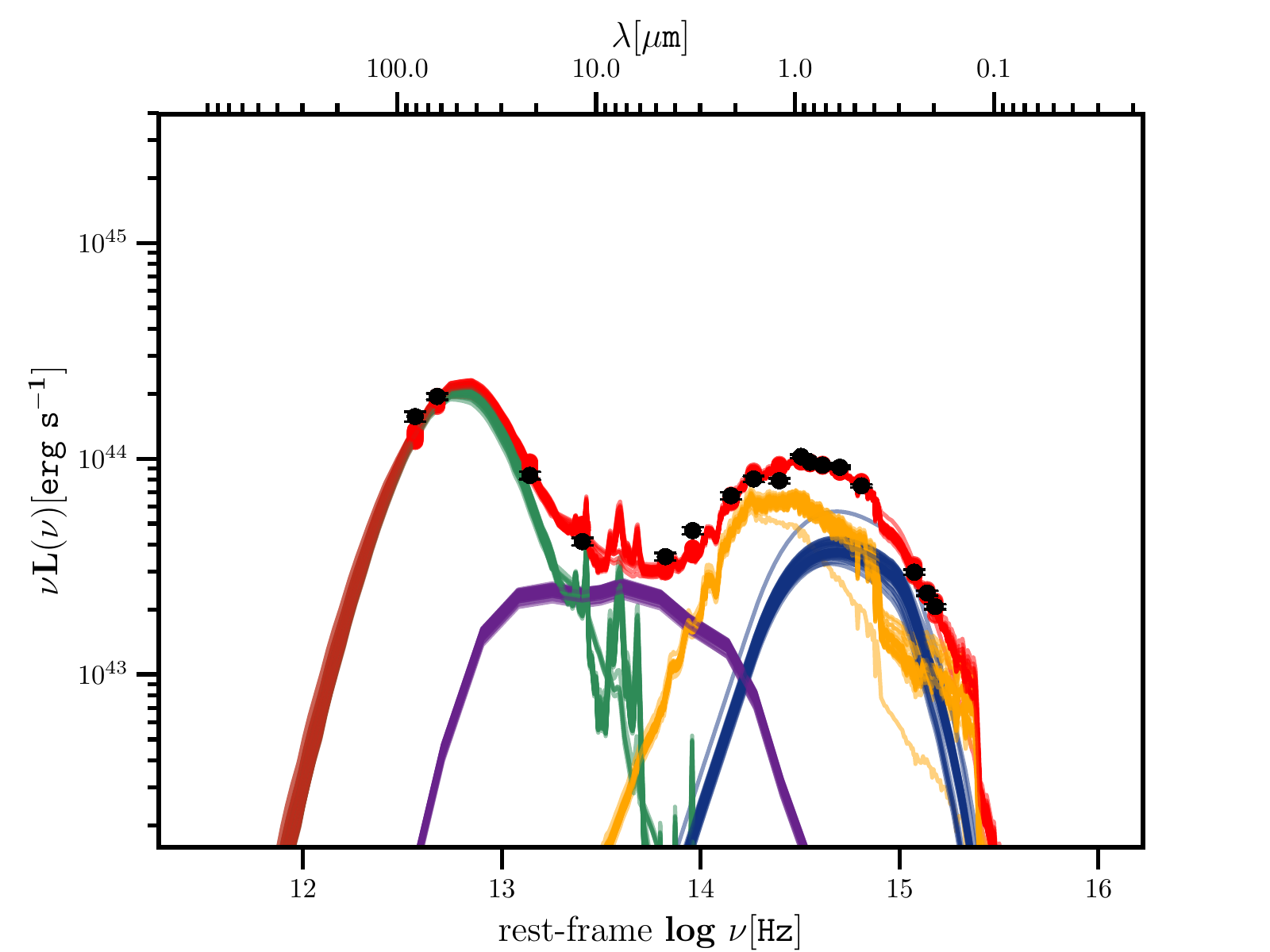}    
    \caption{Same as Fig.~\ref{fig:SED_HE0433} for HE~1108$-$2813.}
    \label{fig:SED_HE1108}
\end{figure}

\begin{figure}
    \includegraphics[width=1.\linewidth]{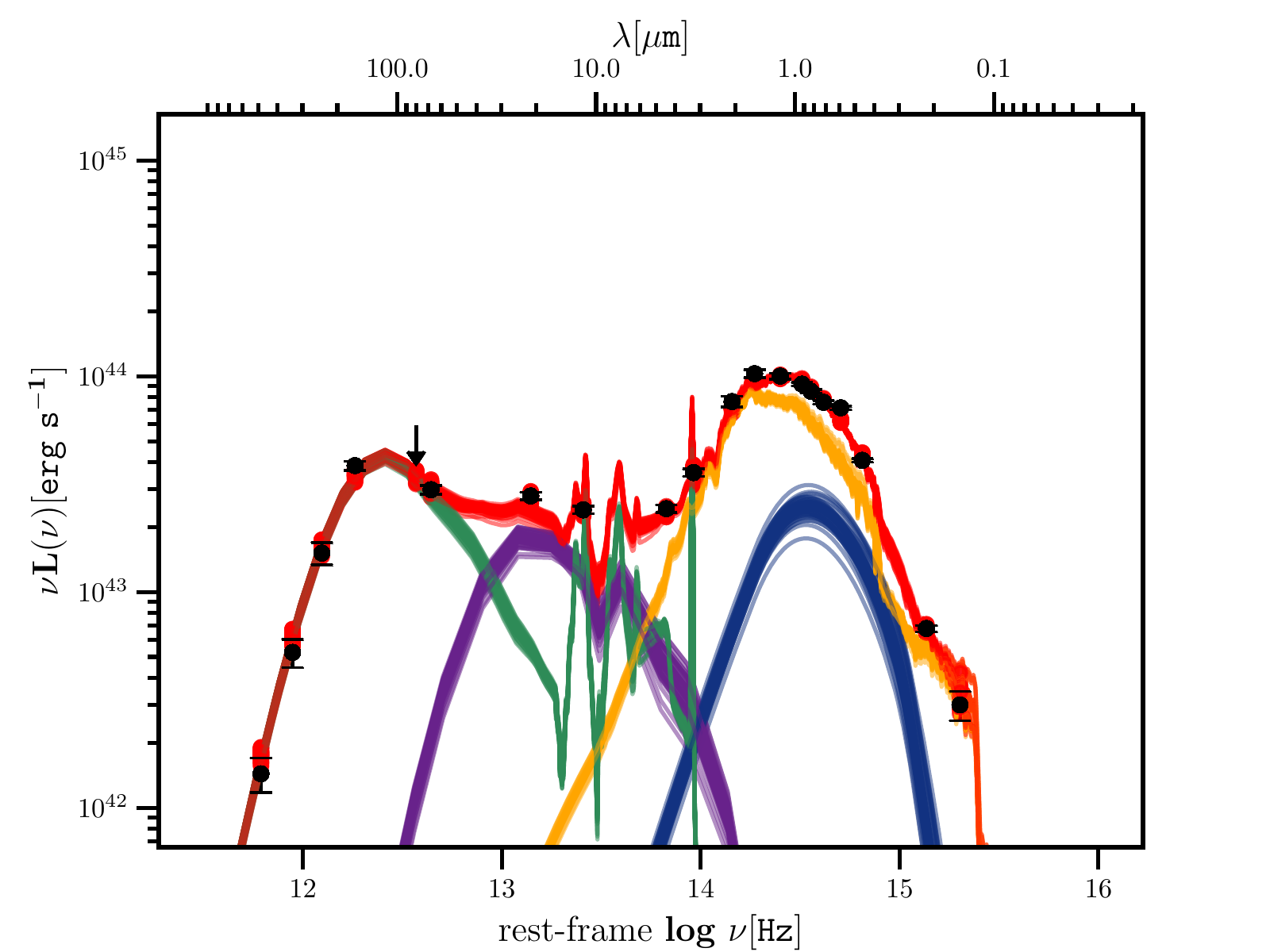}    
    \caption{Same as Fig.~\ref{fig:SED_HE0433} for HE~1353$-$1917.}
    \label{fig:SED_HE1353}
\end{figure}

\begin{figure}
    \includegraphics[width=1.\linewidth]{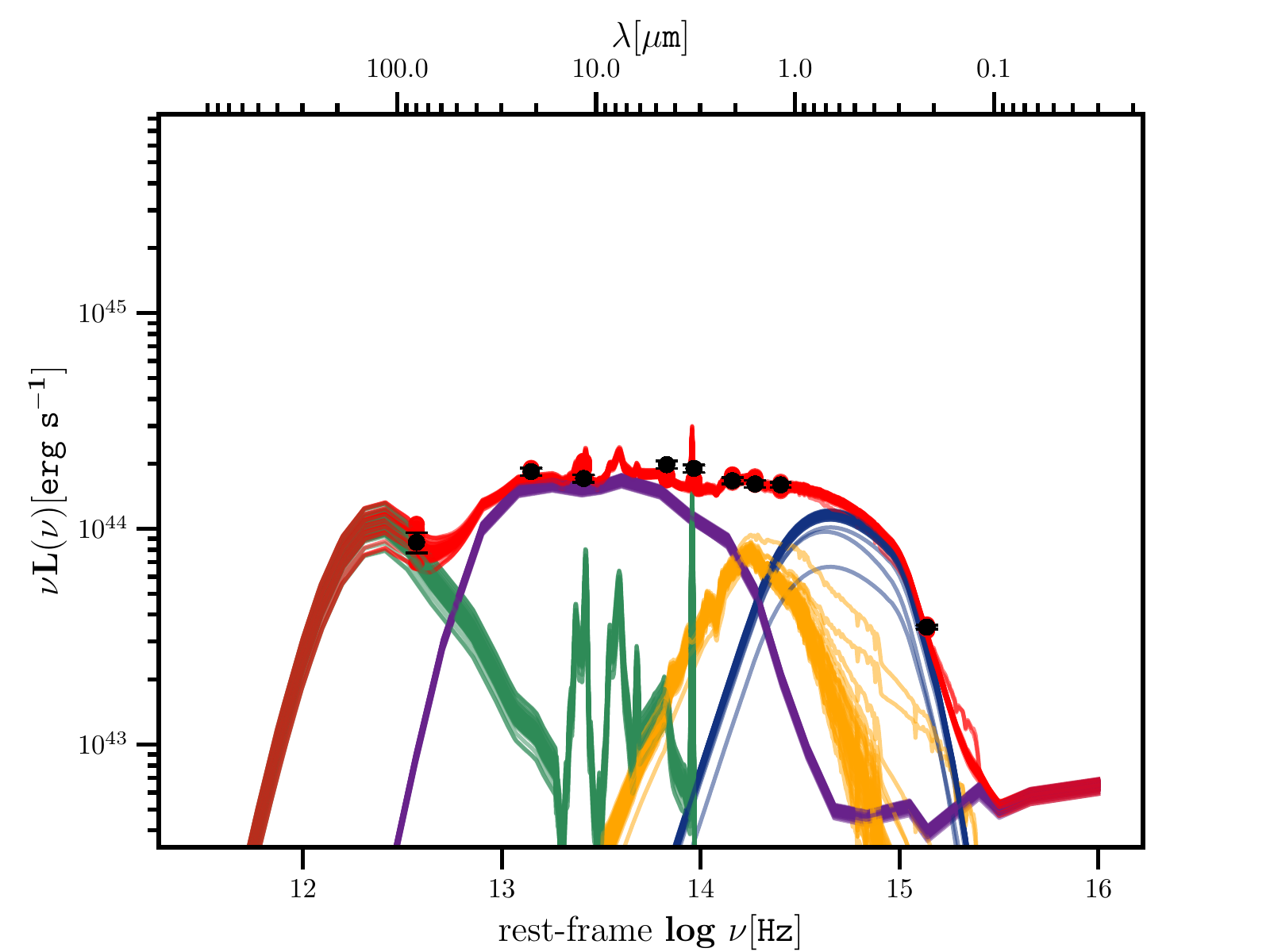}    
    \caption{Same as Fig.~\ref{fig:SED_HE0433} for HE~2211$-$3903.}
    \label{fig:SED_HE2211}
\end{figure}

As described in section~\ref{sec:archival_data}, we used \textsc{AGNfitter} to analyze the observed galaxies. The SED plots are presented in Fig.~\ref{fig:SED_HE0433}--\ref{fig:SED_HE2211}.

\end{appendix}

\end{document}